# The infochemical core

Antoni Hernández-Fernández[1,2,*], Ramon Ferrer-i-Cancho[1]

(1) Complexity and Quantitative Linguistics Lab, LARCA Research Group. Departament de Ciències de la Computació, Universitat Politècnica de Catalunya. Barcelona (Catalonia), Spain.

(2) Institut de Ciències de l'Educació, Universitat Politècnica de Catalunya. Barcelona (Catalonia), Spain.

*Author for correspondence (antonio.hernandez@upc.edu).

*Address correspondence to: Antoni Hernández-Fernández, Institut de Ciències de l'Educació, Universitat Politècnica de Catalunya, Campus Nord, Edifici Vèrtex, Plaça Eusebi Güell, 6, 08034 Barcelona (Catalonia), Spain. Tel: +34 649 00 51 53. Email: antonio.hernandez@upc.edu

**ABSTRACT**

Vocalizations, and less often gestures, have been the object of linguistic research for decades. However, the development of a general theory of communication with human language as a particular case requires a clear understanding of the organization of communication through other means. Infochemicals are chemical compounds that carry information and are employed by small organisms that cannot emit acoustic signals of an optimal frequency to achieve successful communication. Here, we investigate the distribution of infochemicals across species when they are ranked by their degree or the number of species with which they are associated (because they produce them or are sensitive to them). We evaluate the quality of the fit of different functions to the dependency between degree and rank by means of a penalty for the number of parameters of the function. Surprisingly, a double Zipf (a Zipf distribution with two regimes, each with a different exponent) is the model yielding the best fit although it is the function with the largest number of parameters. This suggests that the worldwide repertoire of infochemicals contains a core which is shared by many species and is reminiscent of the core vocabularies found for human language in dictionaries or large corpora.

**1. INTRODUCTION**

Quantitative linguistics is a discipline with a tremendous capacity to explore connections between human language and other natural systems. The key is that a hypothetical connection between language and certain natural systems can be investigated simply by looking at certain statistical properties for which only minimal assumptions are required. For instance, the tendency of more frequent words to be shorter (Zipf, 1935, 1949), is also found in DNA sequences (Naranan & Balasubrahmanyan, 2000; Ferrer-i-Cancho et al., 2013a) and the behaviour of non-human species (see Ferrer-i-Cancho et al., 2013b for a review). A non-trivial version of Menzerath's law, i.e. the tendency of the size of the parts of a linguistic construct to decrease as its number of parts increases (Menzerath, 1954; Altmann, 1980), is found in genomes at different levels of analysis (Wilde & Schwibbe, 1989; Li, 2012; Ferrer-i-Cancho et al., 2013a). In general, the depth of a connection of the kind described above depends on various factors. One factor is the existence of conceptual similarities between both at some level of abstraction (Ferrer-i-Cancho et al., 2013a). For instance, the striking conceptual similarities between human words and codons (Bel-Enguix & Giménez-López, 2011) are reinforced by other similarities arising from statistical analyses (Naranan & Balasubrahmanyan, 2000; Balasubrahmanyan & Naranan, 2000; Naranan, 2011).

Another factor is the number of statistical properties that coincide simultaneously: the likelihood of a non-trivial connection cannot decrease as the number of shared statistical regularities increases (Rao et al., 2012; Ferrer-i-Cancho & McCowan, 2012). For instance, the various statistical similarities between dolphin whistles and human words (McCowan et al., 1999; Ferrer-i-Cancho & McCowan, 2009; Ferrer-i-Cancho & McCowan, 2012) suggest that dolphin whistles may have a communicative function resembling that of human words. However, this is not the only reason why quantitative linguistics has a tremendous potential to bridge the gap between different fields. Due to its minimal assumptions, quantitative linguistics is in a privileged position for selecting candidates for extraterrestrial forms of intelligence (Doyle et al., 2011) or shedding light on the linguistic nature of undeciphered scripts (Rao, 2010).

Contrary to what many believe, quantitative linguistics is more than data analysis or the art of collecting statistical curiosities. A less well-known contribution of quantitative linguistics going back at least to the foundation of modern quantitative linguistics by G. K. Zipf during the first half of the twentieth century (Zipf, 1949), are abstract principles that can explain the recurrence of certain statistical patterns, not only in language but beyond, to ultimately establish laws of language or human behaviour. G. K. Zipf's idea of a minimum equation defining the cost of a set of tools has recently been interpreted as a precursor of the notion of mean code length in information theory (Ferrer-i-Cancho et al., 2013b). Compression, i.e. the minimization of that mean code length has been used to shed light on the origins of the law of brevity, not only in human language but in other species too (Ferrer-i-Cancho et al., 2013b), and DNA sequences (Naranan & Balasubrahmanyan, 2000). A compromise between entropy minimization and mutual information minimization, inspired by Zipf's conflict between unification and diversification (Zipf, 1949), has been used to explain the recurrence of "Zipf's law" patterning in languages (Ferrer-i-Cancho, 2005; Prokopenko et al., 2010; Dickman et al., 2011). This explanation (a) is abstract enough to be valid for DNA (Ji, 1999) and other natural systems where Zipf's-law-like patterning has been found (e.g. Searls, 2002; McCowan et al., 1999) and (b) challenges the view that the solutions of information theory do not resemble those of natural languages (Christiansen & Chater, 2015).

Human vocalizations, in the form of speech or written language (Akmajian et al., 1997), and less intensively gestures (Sandler & Lillo-Martin, 2006), have been objects of research from standard linguistics and neighbouring disciplines for decades. However, the development of a general theory of communication with human language as a particular case requires a clear understanding of the organization of information transfer through other means. While language is believed to be uniquely human (Hauser et al., 2002; Pinker, 2003), a statistically rigorous test of the statement is lacking. While researchers struggle to date the origins of language in modern humans or ancestors, bacteria started linguistic communication millions and millions of years before a vocalizing multicellular organism appeared on Earth (Ben Jacob et al., 2004). Thus, chemical communication might be the oldest communication system produced in our planet. The same applies to writing, a milestone in the development of communicative skills (in humans), being as it is a persistent method of communication that helps one to detach oneself from the here and now. However, bacteria pioneered the exchange of documents, i.e. plasmids, on our planet (Head, 1999/2000).

Language research is anthropocentric: it is based on the assumption that putting human language and human biology at the centre provides a sufficient level of abstraction for solving the puzzle of the origins of language (Hurford, 2012; Fitch, 2010). However, this human-centred vision has a certain flexibility: it allows one to include existing or extinct species that are close in phylogeny or to jump further to species exhibiting, like ourselves, complex vocal behaviour such as songs. This view is challenged again by the non-vocal and non-gestural linguistic communication of brainless unicellular organisms through chemical compounds: bacteria (Ben Jacob et al., 2004). The challenge adds further support for the proposal of a new paradigm for language research including biology and computer science (Bel-Enguix & Jiménez-López, 2012).

Here we consider the particular case of chemical communication in multicellular organisms, a domain that has received, to our knowledge, little attention by quantitative linguistics research, with some exceptions (Doyle, 2009). The goal of this article is applying concepts and tools from quantitative linguistics to shed light on the organization of infochemicals, chemical compounds that carry information (Wyatt, 2003).

**2. INFOCHEMICALS**

The complexity of chemical communication probably requires a zoosemiotic approach (Maran, Martinelli & Turovski, 2011; Riba, 1990), but at present we must settle for analysing signals that can be detected and describe the communicative contexts in which they are emitted. Chemicals provide information (cues) or act as precursors of communication (signals) or as central elements of communication systems that probably evolved from non-communicative compounds with a phylogenetic pattern (Symonds & Elgar, 2004; Steiger et al., 2011, for a review). Chemical signals constitute much of the language of life in the sea (Hay, 2009) and also on dry land (Wyatt, 2003). We humans have a poor capacity to understand chemical interactions, partly due to our rather stunted sense of smell, but our modern technology allows us to explore the fascinating world of infochemicals. Infochemicals are usually divided into two groups: pheromones and allelochemicals (Wyatt, 2010; Dicke & Sabelis, 1988; Nordlund & Lewis, 1976). Pheromones are defined classically as semiochemicals involved in intraspecific communication (Law & Regnier, 1971; Regnier & Law, 1968), substances secreted to the outside by an organism (in contrast with hormones, secreted inside an organism) and perceived by a second individual of the same species in which they release a specific reaction (Karlson & Lüscher, 1959; Wyatt, 2003), in opposition to allelochemicals, which mediate interspecies communications (Nordlund & Lewis, 1976) as happens between plants and insects.

The modern division of infochemicals by function (Wyatt, 2010, for a general review) distinguishes various classes of infochemicals, more or less appropriate according to the area of research concerned (Dicke & Sabelis, 1988). Thus, for example, the classification chosen by El-Sayed for the Pherobase, a free database of infochemicals (El-Sayed, 2012) distinguishes between five behavioural functions (for alternative classifications see Nordlund & Lewis, 1976; Dicke & Sabelis, 1988; Wyatt, 2010): pheromones, involved in intraspecific communication; attractants, or infochemicals that cause aggregation of individuals, secreted by species or synthesized by humans; allomones, or allelochemicals that benefit the sender; kairomones or allelochemicals beneficial for the receiver; and finally synomones, which are allelochemicals benefiting both signaller and receiver in mutualistic interactions.

The Pherobase (El-Sayed, 2012) is a wide-ranging database that incorporates a list of species that produce, or are sensitive to, each infochemical, as well as other biochemical characteristics. Pheromones and allelochemicals are a way of transmitting information whose main advantage is their ability to spread and persist in the environment in which they expand. Their diffusion is conditioned to the predominant species' habitat and its activity (Okubo et al., 2001), as had already been concluded by pioneer studies of infochemicals (Wilson, 1958; Wilson & Bossert, 1963; Wilson, 1970).

***TABLE 1 NEAR HERE***

Here, the degree of an infochemical is defined as the number of species that are associated with it, because they produce it or are sensitive to it, according to the Pherobase (El-Sayed, 2012). The infochemical with the highest degree has rank 1, the second has rank 2, and so on (Table 1). The number of functions that an infochemical serves for a given species is irrelevant for our notion of degree. For instance, if an infochemical is associated with only one species, the infochemical will have degree one regardless of the number of functions served.

In this article, the fit of different functions to the rank distribution of infochemical associations in nature is studied. The list of functions considered is summarized in Table 2. It is found that the function providing the best balance between the exactness of the fit and the number of parameters is a double Zipf (a power law with two different exponents) although it is the function with the largest number of parameters. This suggests that infochemicals have a core repertoire analogous to the core vocabulary found in human language (Ferrer-i-Cancho & Solé, 2001; Petersen et al., 2012; Gerlach & Altmann, 2013; Cocho et al., 2015).

***TABLE 2 NEAR HERE***

## 3. MATERIALS AND METHODS

*f(r)* is defined as the degree of rank *r*, *n* as the maximum rank (*r* = 1,2,…,*n*) and *T* as the sum of all the degrees, i.e.

$$T = \sum_{r=1}^{n} f(r), \tag{1}$$

where *n* is the size of the repertoire.

Various two-parameter models such as Zipf's function, Beta function, Yule function or Menzerath-Altmann function (see Table 2) can be fitted to rank-frequency data (Li et al., 2010). Here, Li et al.'s (2010) model-selection methodology is adopted to study rank-degree data in infochemicals. A two regime distribution Zipf distribution (Ferrer i Cancho & Solé, 2001) is added to the list of functions explored by Li et al. (2010).

### 3.1. Materials

The degree and the classification of each infochemical come from the Pherobase (El-Sayed 2012) which is freely available at http://www.pherobase.com/. A study primarily concerned with compounds which exist in nature and which regulate communication without human intervention should discard attractants synthesized in human laboratories. Therefore, two kinds of analyses are considered: one concentrating on the pheromones and allelochemicals of the Pherobase (El-Sayed, 2012) and another on the whole database, including attractants synthesized by humans. The whole database comprises a repertoire of $n$=1686 chemical compounds and $T$=17633 species-infochemical associations. A summary of the elementary features of the more frequent elements of the dataset is shown in Table 1.

### 3.2. Methods

The methodology for fitting functions to the dependency between rank and degree is borrowed from Li et al.'s for the dependency between the rank of a word and its frequency (Li et al., 2010). Baayen's (2008) methodology is used to compute the breakpoint parameter of the double Zipf (parameter $r^*$ in Table 2). Li et al.'s (2010) methodology consists of a linear regression of the target function on a double logarithmic transformation, i.e. $\log(r)$ versus $\log(f_r)$, and then using Akaike's information criterion, a combination of likelihood to evaluate the quality of the fit and a penalty for the number of parameters used. Table 2 summarizes the functions that are fitted to the rank distribution of infochemicals and the corresponding double logarithmic transformation that is used for linear regression.

$SSE$ is defined as the sum of the squared differences between the logarithm of $f_r$, the observed degree of rank $r$, and the logarithm of $F_r$, expected degree of rank $r$, i.e.

$$SSE = \sum_{r=1}^{n} w_r (\log(f_r) - \log(F_r))^2 \tag{2}$$

$w$ being a weight that is $w_r = 1$ in unweighted regression and $w_r = 1/r$ for weighted regression following Li et al.'s (2010) methodology. The goal of the weighted regression is to give more importance to low ranks.

The log-likelihood *L* is defined (Venables & Ripley, 1999; Li et al., 2010) as

$$L = C - \frac{n}{2}\log\left(\frac{SSE}{n}\right) - \frac{n}{2} \tag{3}$$

where *n* is the maximum rank and *C* is an additive constant that depends on the model (see Table 2). For model selection, the Akaike Information Criterion (AIC) is used as in Li et al. (2010). As Li et al. (2010) point out, any constant term of Eq. 3 will be cancelled when two models are compared and then the AIC defined by Akaike (1974) is:

$$AIC = -2L + 2K = n\log\left(\frac{SSE}{n}\right) + 2K, \tag{4}$$

where *K* is the number of free parameters (Table 2) in the model under consideration.

The AIC difference of a model, Δ, is defined as the difference between the AIC of the model and that of the model with smallest AIC (Li et al., 2010). Then, trivially Δ=0 for the best model. The relationship between AIC and SSE defined in equation 3 is still true for weighted regression as Li et al. (2010) demonstrate, but the correlation-based regression of $R^2$ for weighted regression in logarithmic scale has a specific definition (see Li et al. (2010) for further details).

To fit the double Zipf, a function not considered by Li et al. (2010), Baayen's (2008, pp.234-239) technique is used to compute the breakpoint automatically. That breakpoint is the rank *r\** that defines the boundary between two consecutive simple Zipf distributions. The optimal breakpoint *r\** is the rank that minimizes the *SSE*, which is obtained by exhaustively exploring all the possible values of *r\**, that is fitting the double Zipf with a given *r\** that varies between 1 and *n*, and keeping the *r\** yielding the smallest *SSE*.

\*\*\*FIGURE 1 NEAR HERE\*\*\*

Fig. 1 indicates that the breakpoint obtained for the Pherobase is not a local optimum. The R statistical package is used to analyse all the data (see R online manuals at http://www.r-project.org/ and Baayen (2008)). Table 2 summarizes the list of functions fitted to the rank distribution of degree.

**4. RESULTS**

Table 3 and Table 4 summarize the fitting of different functions to the empirical relationship between the degree of an infochemical and its rank according to the Pherobase (El-Sayed 2012), including or excluding attractants, respectively. Both the correlation coefficient ($\rho$) and AIC differences ($\Delta$) suggest that the double Zipf is the function providing the best fit, both in the weighted and unweighted regression, regardless of whether attractants are included or not. All the other functions considered (simple Zipf, Beta, Yule and Menzerath-Altmann) are far from yielding the best fit: the second best function, both in unweighted and weighted regression for the whole database, is Yule function with AIC differences $\Delta$=967.5 and $\Delta$=1450, respectively (Table 3) and similar results for the analysis excluding attractants (Table 4). AIC differences of the order of a thousand are normally considered sufficient to discard the model under consideration (Burham & Anderson, 2002).

\*\*\*TABLE 3 AND TABLE 4 NEAR HERE\*\*\*

Figure 2 shows the best fit of the double Zipf equation, with and without weights, for the whole database (Fig 2 A and C) and without attractants (Fig 2 B and D), respectively. When all infochemicals are considered, the breakpoint is $r^*$= 180 for unweighted, and $r^*$=79 for weighted regression. When attractants are excluded, the breakpoint is $r^*$=151 for unweighted, and $r^*$=61 for weighted regression (see Table 5). The value of the breakpoint parameter of the double Zipf suggests that infochemicals are divided into two groups: a core of the order of one hundred infochemicals and the remainder.

***TABLE 5 AND FIGURE 2 NEAR HERE***

## 5. DISCUSSION

This double Zipf distribution of ranks (in the degree of an infochemical as a function of its rank rank) is also found in the rank distribution of words, i.e. the frequency of occurrence of a word (in tokens) as a function of its rank (Ferrer-i-Cancho & Solé, 2001; Petersen et al., 2012; Gerlach & Altmann, 2013; Cocho et al., 2015), where the breakpoint defines the boundary between a vocabulary core, i.e. a finite vocabulary of semantically versatile word types, and a potentially infinite peripheral vocabulary.

The connection may not be obvious at first glance: our target has not been the frequency of occurrence of an infochemical in nature but its degree, namely, the number of species that produce it or are sensitive to each infochemical. However, word frequency is connected lawfully with another linguistic variable, namely word polytextuality, which can be defined as the number of texts in a corpus where the target word appears at least once (Köhler, 1986). The metaphor that words types are infochemicals and texts are the infochemicals from the environment that the members of a species have produced or been sensitive to, and the positive correlation between frequency and polytextuality (Köhler, 1986), suggest that the infochemicals in the high degree regime (the low ranks) form a core repertoire analogous to the core lexicon found in the high frequency regime of human language (Ferrer-i-Cancho & Solé, 2001), while the infochemicals in the low degree regime would form a peripheral repertoire analogous to the peripheral lexicon found in the lower frequency domain of words (Ferrer-i-Cancho & Solé, 2001).

Cores have also been investigated in cognitive networks (Baronchelli et al., 2013). A network core is defined by Baronchelli et al. (2013) as "*a powerful subset of the network because of the high frequency of occurrence of its nodes, their importance for the existence of remainder of nodes, or the fact that it is both densely connected and central (in terms of graph distance)*". A network analysis of cross-referencing between dictionary entries has shown that dictionaries have a core consisting of about 10% of words (typically with a concrete meaning and acquired early), from which other words can be defined (Picard et al., 2009). In our case, we are analysing a bipartite graph of infochemical-species associations (one partition for infochemical and another partition for species). Table 5 summarizes the proportion of infochemical types within the chemical core taking the breakpoint of the double Zipf as the boundary. The percentage of infochemicals that belong to the core repertoire is about 10% according to unweighted regression, about the same percentage of word types in the grounding kernel of a dictionary (Picard et al., 2009). The 10% of the core both in dictionaries and infochemicals might be simply anecdotal and should be explored further. The percentage of infochemical-species associations within the chemical kernel is two-thirds (11755 infochemicals in the core repertoire over 17633 in total, just 66.66%) of the total, but this percentage varies according to the kind of function and increases to more than seventy percent for attractants and allomones (Table 6). Similarly, the core vocabulary of words identified by means of a double Zipf by Ferrer-i-Cancho & Solé (2001) is responsible for 85% of word tokens in a large English multiauthor corpus. Although infochemical degrees and word frequencies are not fully comparable, the high percentages of associations/tokens in the core highlights the importance of cores.

***TABLE 6 NEAR HERE***

Pheromones and kairomones are just a little below sixty percent of infochemical associations in the chemical kernel (Figure 3) and both have a greater presence than attractants and allomones (in percentage) in the peripheral chemical repertoire, with synomones not appearing because their number is very low (Figure 3). The presence of pheromones increases in the peripheral chemical repertoire probably as a response to a major necessity of communication specificity and to reduce communication interferences with other species. The information carried by general attractants of the infochemical core may need to be completed and detailed by peripheral, perhaps species-specific, pheromones to avoid confusion.

***FIGURE 3 NEAR HERE***

The chemical kernel constitutes a core of infochemicals shared by many species (Table 5). We hypothesize that the design of such a core could be driven by the economy of its compounds from the perspective of the sender (e.g. ease of production of the compound) and communicative efficiency in a given environment from the perspective of the receiver. Communicative efficiency is determined by various factors such as ease of detection or persistence of the chemical compound. The heterogeneity of the ecosystems of the species included in the database probably limits the cases of communicative interference between species. Furthermore, to avoid the confusion that might arise in an ecosystem with different species using the same infochemicals, species may adopt different diffusion strategies and exploit the zoosemiotic communication context to reduce confusion (Maran, Martinelli & Turovsky, 2011; Riba, 1990).

The most frequently analyzed infochemicals (Table 2), by definition on the core chemical repertoire, include, for instance, some isomers of tetradecenyl acetate that are well-known sex pheromones (Shorey, 1976; Chapman, 1998) and are attractive to all of the males in a large group of different closely-related species. As already pointed out by Shorey (1976) in his classical review, other chemicals present in relatively small quantities enhance the attractiveness of the pheromone for males of the correct species and, at the same time, inhibit attraction of males of wrong species.

In fact, each species has its own chemical communication system, with a finite set of associated infochemicals that it can emit, $V_e$, and another set to which it responds, $V_r$. Actually, $V_e \neq V_r$ since there are semiochemicals (e.g., synthesized compounds) that a species will not emit, but will be attracted to. Using set theory, the chemicals that constitute a chemical communication system is $V_e \cup V_r$, while the elements emitted and also detectable are $V_e \cap V_r$. In this Universe of chemical communication there are species that are sensitive to a shared chemical repertoire, emitted by themselves or not. It is a complex Universe because there are substances that, within the same species, can attract females and not males or vice versa.

The analysis of non-human communication systems may be able to provide insights into the efficiency of signalling systems that might otherwise be inaccessible (Doyle, 2009). Keeping a certain distance, parallels between our study and other linguistic phenomena, e.g. shared elements between different languages, can be established. Above we have considered the metaphor of a species as a text of related infochemicals. Here we regard the infochemicals with which a species is related as a language. There are infochemicals shared in the communication systems of different species thus, arguably, species may speak "different chemical languages" but do have some communicative elements in common (not necessarily with the same meaning). Similarly, different human languages can share word elements such as phonemes or syllables in their linguistic systems. Some languages not only share those building blocks of words but also word themselves. In a more complex way, languages sometimes share elements like words of basic vocabulary, e.g. a Swadesh list, especially if they are related languages. From an evolutionary perspective, related languages tend to maintain diachronically basic words in the same or practically identical word form (Wang & Wang, 2004) just as different species can be sensitive to very similar —or the same— semiochemicals when they come from the same phylum (Symonds & Elgar, 2004).

The ubiquity and variety of pheromones can be explained by natural selection (Wyatt, 2009; Hauser, 1996). From a Zipfian perspective (Zipf, 1949), the existence of a peripheral chemical repertoire (the infochemicals outside of the core) could be a consequence of the need to diversify species' communication possibilities, which would be particularly useful in a noisy channel. If the principle of least effort leads all species to emit similar chemicals compounds from the core because of their high availability, ease of production or utility in different environments (Okubo et al., 2001), then interference and confusion emerges as inevitable, especially in very rich terrestrial ecosystems like tropical forests (Basset et al., 2012). Thus, diversification is opposed to unification (Ferrer-i-Cancho & Solé, 2003; Zipf, 1949) also in chemical communication systems. From a Darwinian evolution perspective, it is expected that an optimal occupation of the chemical channel in ecosystems arises over time. In sum, the existence of both a chemical core and a periphery could be a natural consequence that communication has to solve two different problems: efficient coding of information and transmission (Ferrer-i-Cancho et al., 2013).

Although linguistics is mainly concerned with the vocal modality, interest in gestural language through conventional sign language or spontaneous gestures has been increasing over time (Goldin-Meadow et al., 2008; Sandler & Lillo-Martin, 2006). The chemical modality might be the next frontier to be explored intensively. Quantitative linguistics now has the challenge of understanding the mechanisms underlying the emergence of two regimes in words and infochemicals and identifying the distinctive features of those cores (Gerlach & Altmann, 2013; Cocho et al., 2015). Throughout this long research, we hope to answer a very important question: what are the principles of organization shared between human language and chemical communication? Understanding both as evolutionary and complex adaptative systems might be crucial (Beckner et al., 2009; Bel-Enguix & Jiménez-López, 2012).

The quantitative exploration of the Pherobase is just beginning and our analysis has focused on the large scale. Future research should pay attention to specific ecological niches or a concrete phylum. The generosity of those who share data on infochemicals is helping to explore connections between apparently distant domains, which will become increasingly more common in twenty-first century science.

ACKNOWLEDGEMENTS

We thank G. Bel-Enguix for helpful comments and discussions. We are grateful to A.M. El-Sayed and all the people who have made the Pherobase® possible and to Jordi Las for his help with the programming of the data acquisition. RFC and AHF wish to express their gratitude for support in the form of (a) the grant 2014SGR 335 890 (MACDA) from AGAUR (Generalitat de Catalunya) and (b) the APCOM project (TIN2014-57226-P) from MINECO (Ministerio de Economia y Competitividad). Additionally, RFC is grateful to the Universitat Politècnica de Catalunya for support in the form of the grant "*Iniciació i reincorporació a la recerca*" and to the Spanish Ministry of Science and Innovation for the grants BASMATI (TIN2011-27479-C04-03) and OpenMT-2 (TIN2009-14675-C03).

**Table 1:** List of the ten infochemicals with the highest degree (in number of species) according to the Pherobase (El-Sayed 2012). $f_r$ is the degree of the infochemical with the $r$-th largest degree. Each chemical compound can serve different biological functions.

| Infochemicals | r | $f_r$ |
|---|---|---|
| (Z)-9-Tetradecenyl acetate | 1 | 342 |
| (Z)-11-Tetradecenyl acetate | 2 | 309 |
| (Z)-11-Hexadecenyl acetate | 3 | 278 |
| 2,6,6-Trimethylbicyclo[3,1,1]hept-2-ene | 4 | 245 |
| (Z)-7-Dodecenyl acetate | 5 | 235 |
| (E)-11-Tetradecenyl acetate | 6 | 229 |
| Carbon dioxide | 7 | 222 |
| Ethanol | 8 | 221 |
| (Z)-11-Hexadecenal | 9 | 185 |
| 1-Octen-3-ol | 10 | 183 |

**Table 2:** List of functions used in this article (simplified and adapted from Li et al., 2010 and enriched with the double Zipf) and the parameters fitted, i.e. *a*, *b* and normalization constants (*C* and *C'*). In the double logarithmic transformation for linear regression, $C_0$ is the independent term and $C_1$ is the coefficient that multiplies log *r*. *r\** is the rank of the breakpoint in the double Zipf function and defines two regimes, a 1st regime from *r*=1 to *r*=*r\** and a 2nd regime for *r*>*r\**.

| MODEL | FREE PARAMETERS (*K*) | ORIGINAL FUNCTION (*r*=rank, $F_r$=degree, *n*=repertoire size) | DOUBLE LOGARITHMIC TRANSFORMATION |
|---|---|---|---|
| Zipf | 1 | $F_r = \dfrac{C}{r^a}$ | log $F_r$ = $C_0$ + $C_1$ log *r* |
| Beta | 2 | $F_r = \dfrac{C(n+1-r)^b}{r^a}$ | log $F_r$ = $C_0$ + $C_1$ log *r* +$C_2$log(*n*+1-*r*) |
| Yule | 2 | $F_r = \dfrac{C \cdot b^r}{r^a}$ | log $F_r$ = $C_0$ + $C_1$ log *r* +$C_4$*r* |
| Menzerath-Altmann | 2 | $F_r = C \cdot r^b \cdot e^{-a/r}$ | log $F_r$ = $C_0$ + $C_1$ log *r* + $C_3$/*r* |
| Double Zipf | 3 | 1st regime: $F_r = \dfrac{C}{r^a}$ ; 2nd regime: $F_r = \dfrac{C'}{r^{a'}}$ | 1st regime: log $F_r$ = $C_{01}$ + $C_{11}$ log *r*<br>2nd regime: log $F_r$ = $C_{02}$ + $C_{12}$ log *r* |

**Figure 1:** The sum of squared errors of the double Zipf, *SSE*, versus the breakpoint rank, $r^*$. A and B are the plots for the whole Pherobase, unweighted and weighted respectively. C and D are analogous to A and B excluding attractants synthesized by humans, unweighted and weighted, respectively. It can be seen that the breakpoint $r^*$ obtained corresponds to a global minimum of deviance in all cases. Only in C do we find another local minimum for very large ranks that is not relevant due to its comparatively large *SSE* value with regard to the global minimum.

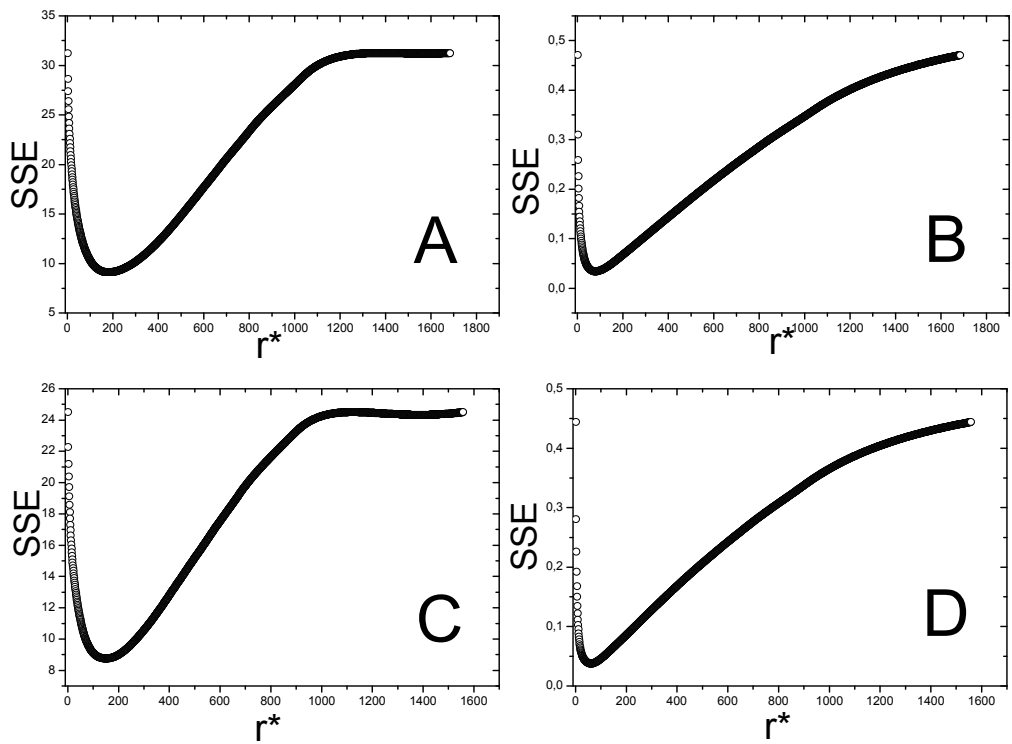

**Figure 2:** Degree (number of species that that are associated with each infochemical) versus the infochemical rank in double logarithmic scale (white circles) versus the best fit of the double Zipf model (solid line). A and B correspond to the whole database while attractants are excluded for C and D. A and C are unweighted while B and D are weighted. In each subplot, the rank breakpoint ($r^*$), the Zipf's exponent for the first and second regime ($C_{11}$ and $C_{12}$, respectively) are shown.

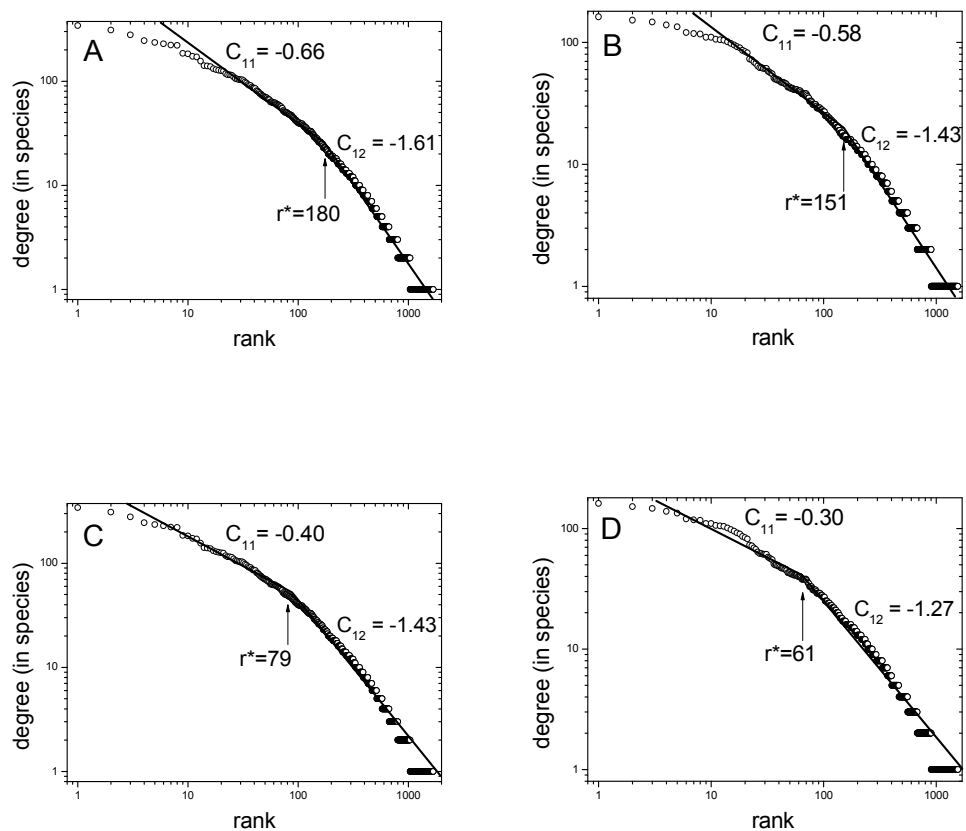

**Table 3:** Summary of the results of the fitting of the ensemble of functions to the whole database (El-Sayed, 2012). For every target function, the coefficients giving the best fit are shown (the meaning of each coefficient is explained in Table 1). $\Delta$ is the difference between the AIC of the target function and that of the function giving the lowest AIC), *SSE* is the sum of squared errors and $\rho$ is the correlation coefficient ($R^2_{corr,log}$ and $R^2_{corr,log,weight}$, for unweighted and weighted fit, respectively, following Li et al.'s (2010) notation).

| Function | Unweighted | | | | Weighted | | | |
|---|---|---|---|---|---|---|---|---|
| | Coefficients | $\Delta$ | *SSE* | $\rho$ | Coefficients | $\Delta$ | *SSE* | $\rho$ |
| Zipf | $C_0=4.12\pm 0.02$<br>$C_1=-1.29\pm0.01$ | 2074.4 | 31.24 | 0.94 | $C_0=2.84\pm 0.01$<br>$C_1=-0.74\pm0.01$ | 4425.2 | 0.47 | 0.90 |
| Beta | $C_0=3.59\pm 0.05$<br>$C_1=-1.21\pm 0.01$<br>$C_2=0.051\pm0.004$ | 1939.9 | 28.81 | 0.95 | $C_0=-0.17\pm 0.08$<br>$C_1=-0.625\pm0.005$<br>$C_2=0.39\pm0.01$ | 3426.4 | 0.26 | 0.95 |
| Yule | $C_0=3.36\pm 0.02$<br>$C_1=-0.90\pm 0.01$<br>$C_4=-39\cdot10^{-5}\pm1\cdot10^{-5}$ | 967.5 | 16.18 | 0.97 | $C_0=2.66\pm 0.01$<br>$C_1=-0.49\pm 0.03$<br>$C_4=-92\cdot10^{-5}\pm1\cdot10^{-5}$ | 1450.5 | 0.08 | 0.98 |
| Menzerath-Altmann | $C_0=4.41\pm 0.02$<br>$C_1=-1.39\pm 0.01$<br>$C_3=-2.9\pm0.1$ | 1354.7 | 20.36 | 0.96 | $C_0=3.40\pm 0.01$<br>$C_1=-0.98\pm 0.01$<br>$C_3=-0.95\pm0.02$ | 2756.8 | 0.17 | 0.96 |
| **Double Zipf** | $C_{01}=1.43\pm 0.01$<br>$C_{11}=-0.66\pm 0.01$<br>$C_{12}=-1.61\pm0.01$ | **0** | **9.11** | **0.98** | $C_{01}=1.83\pm 0.01$<br>$C_{11}=-0.40\pm 0.01$<br>$C_{12}=-1.43\pm0.01$ | **0** | **0.034** | **0.99** |

**Table 4:** Summary of the fitting of the ensemble of functions to the Pherobase (El-Sayed, 2012) excluding attractants. The format is the same as in Table 3.

| Function | Unweighted | | | | Weighted | | | |
|---|---|---|---|---|---|---|---|---|
| | Coefficients | $\Delta$ | *SSE* | $\rho$ | Coefficients | $\Delta$ | *SSE* | $\rho$ |
| Zipf | $C_0 = 3.68 \pm 0.02$<br>$C_1 = -1.17 \pm 0.01$ | 1601.5 | 24.50 | 0.94 | $C_0 = 2.52 \pm 0.01$<br>$C_1 = -0.66 \pm 0.01$ | 3845.3 | 0.44 | 0.88 |
| Beta | $C_0 = 3.34 \pm 0.05$<br>$C_1 = -1.12 \pm 0.01$<br>$C_2 = 0.033 \pm 0.004$ | 1543.6 | 23.58 | 0.94 | $C_0 = -0.26 \pm 0.08$<br>$C_1 = -0.55 \pm 0.01$<br>$C_2 = 0.37 \pm 0.01$ | 3030.3 | 0.26 | 0.93 |
| Yule | $C_0 = 3.08 \pm 0.02$<br>$C_1 = -0.85 \pm 0.01$<br>$C_4 = -35 \cdot 10^{-5} \pm 1 \cdot 10^{-5}$ | 881.3 | 15.41 | 0.96 | $C_0 = 2.34 \pm 0.01$<br>$C_1 = -0.42 \pm 0.01$<br>$C_4 = -95 \cdot 10^{-5} \pm 1 \cdot 10^{-5}$ | 1448.2 | 0.1 | 0.98 |
| Menzerath-Altmann | $C_0 = 3.95 \pm 0.02$<br>$C_1 = -1.26 \pm 0.01$<br>$C_3 = -2.6 \pm 0.1$ | 924.9 | 15.85 | 0.96 | $C_0 = 3.09 \pm 0.01$<br>$C_1 = -0.98 \pm 0.01$<br>$C_3 = -0.95 \pm 0.02$ | 2107.5 | 0.15 | 0.96 |
| **Double Zipf** | $C_{01} = 1.32 \pm 0.01$<br>$C_{11} = -0.58 \pm 0.01$<br>$C_{12} = -1.43 \pm 0.01$ | **0** | **8.76** | **0.98** | $C_{01} = 1.73 \pm 0.01$<br>$C_{11} = -0.30 \pm 0.01$<br>$C_{12} = -1.27 \pm 0.01$ | **0** | **0.038** | **0.99** |

**Table 5:** Percentage of infochemical types both in the core and peripheral chemical repertoire considering the breakpoint $r^*$ of the two regime distribution as the boundary between the two (El-Sayed, 2012). $T$ is the number of infochemical-species associations (the total sum of degrees) and $n$ is the repertoire size (in infochemical types).

|  | All infochemicals | | All infochemicals without attractants | |
|---|---|---|---|---|
|  | **Unweighted** | **Weighted** | **Unweighted** | **Weighted** |
| **Core chemical repertoire (1st regime)** | 10.68% | 4.69% | 9.69% | 3.91% |
| **Peripheral chemical repertoire (2nd regime)** | 89.32% | 95.31% | 90.31% | 96.09% |
| **Breakpoint ($r^*$)** | 180 | 79 | 151 | 61 |
| **Total infochemical-species associations ($T$)** | 17633 | 17633 | 11380 | 11380 |
| **Repertoire ($n$)** | 1686 | 1686 | 1560 | 1560 |

**Table 6:** Number of infochemical-species associations that are within the "Core Chemical Repertoire", and in the "Peripheral Chemical Repertoire", by kind of infochemical and for the whole database (percentages are shown in parentheses and are relative to the total in the right-most column). The breakpoint $r*$ of the double Zipf function in unweighted regression defines the boundary between both repertoires.

|  | Attractants | Allomones | Kairomones | Pheromones | Synomones | TOTAL |
|---|---|---|---|---|---|---|
| **Core repertoire** | 4833 (41.11%) | 1506 (12.81%) | 410 (3.49%) | 5005 (42.57%) | 1 (<0.01%) | 11755 (100%) |
| **Peripheral repertoire** | 1420 (24.16%) | 588 (10.00%) | 302 (5.14%) | 3563 (60.62%) | 5 (0.09%) | 5878 (100%) |
| **TOTAL** | 6253 (35.46%) | 2094 (11.88%) | 712 (4.04%) | 8568 (48.59%) | 6 (0.03%) | 17633 (100%) |

30 HERNÁNDEZ-FERNÁNDEZ & FERRER-I-CANCHO

**Figure 3:** Percentage of infochemical associations within the "Core Chemical Repertoire", and in the "Peripheral Chemical Repertoire" by kind of infochemical. Synomones cannot be seen due to their very low proportion. Data borrowed from Table 6.

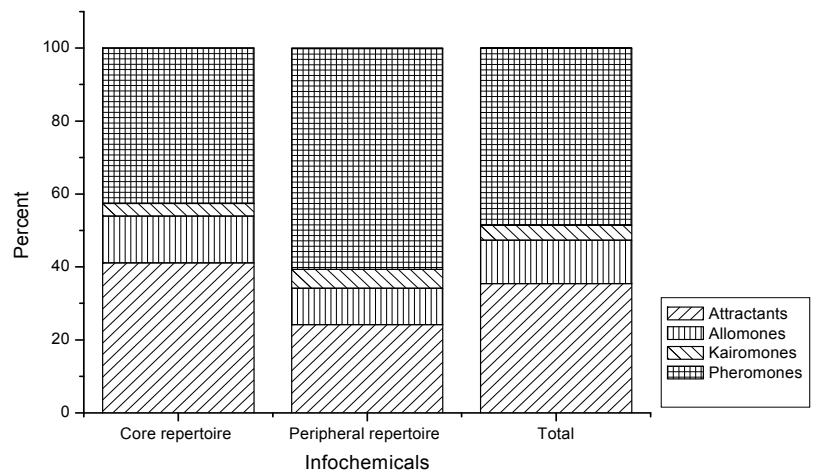